\documentclass[epj]{svjour}
\usepackage{graphicx,hyperref,amsmath,amsfonts,amssymb,widetext,color}

\def\p{\partial}
\def\bn{\textbf{n}}

\def\bi{\textbf{e}}

\def\ve{\varepsilon}
\def\vp{\varphi}
\def\be{\begin{equation}}
\def\ee{\end{equation}}

\def\rev#1{\bgroup  #1 \egroup}

\begin{document}

\title{Shaping thin nematic films with competing boundary conditions}

\author{O V Manyuhina}  

\institute{Nordita, Royal Institute of Technology \& Stockholm University,  Roslagstullsbacken~23, SE-10691 Stockholm, Sweden\\ Present address: Physics Department, Syracuse University, Syracuse, NY 13244, USA}
\date{\today}
%
\abstract{Free interfaces of liquid crystals tend to minimise both capillarity and anchoring forces. 
Here we study nematic films in planar and radial geometries with antagonistic anchoring boundary conditions and one deformable interface. Assuming a perturbation {\it ansatz} we study possible couplings of the director configuration with the shape of free interfaces. In the long-wavelength limit independent of the surface tension, we find analytically threshold thickness when flat film becomes unstable. Next we quantify the bifurcation of a circular ring towards structures with $m$-fold rotational symmetry, induced by elastic anisotropy of nematic director in the bulk.  We believe that our simplified approach can give additional  insight into elastic and capillary phenomena of materials with inherent liquid crystalline order and free interfaces.}

\maketitle
%


\section*{Introduction}

Liquid crystalline matter of synthetic or biological nature is endowed with long-range orientational order, described by the unit vector $\bn$, called the director characterising the averaged orientation of molecules. The presence of interfaces, inclusions or formation of defects can disrupt this order, causing the change of director's orientation and thus elastic deformations on a certain length-scale. Alternatively, if the interface is free, the system can resolve frustration by changing the shape of its interface, whence minimising elastic distortions of the director $\bn$. The interplay between the spatial variation of $\bn$ and the shape of interface requires the integrated multiscale modelling of bulk, surface and contact lines, which is crucial, in particular, for applications of liquid crystal (LC) theory to biological materials~\cite{rey:2007}.

The instability of free interfaces in presence of magnetic fields towards a singular hill-and-valley structure  was predicted by deGennes in 1970~\cite{degennes:1970} and later observed experimentally at the nematic--isotropic interface. This instability results from the competition between elasticity of the director $\bn$, capillarity and gravitational forces~\cite{oswald:2010}. Authors of~\cite{popanita:2003,raghu:1995} has shown that the threshold is also influenced by the boundary conditions, which account for the orientation of $\bn$ with respect to the surface normal, known as anchoring. Indeed,  close to the nematic--isotropic transition temperature, the surface tension $\gamma$ is weak. Therefore the anisotropic anchoring $W_a$ plays an important role in `shaping' interfaces by minimizing elastic distortions of $\bn$ in expense of capillary waves. At the nematic--air interface, on the contrary, the surface tension $\gamma\simeq10^{-2}$~J/m$^2$, which is  several orders of magnitude larger than the anchoring strength $W_a\simeq 10^{-5}$~J/m$^2$. Thus, {\it a priori} one assumes a flat interface yielding the least surface area. Nevertheless, elasticity and anchoring can be driving forces to destabilize and spontaneously deform free interfaces of liquid crystals, similar to Plateau--Rayleigh, Rayleigh--Taylor and Rosenweig surface instabilities in conventional fluids~\cite{book:wetting} triggered by the surface tension, gravity and magnetic fields.

In this paper we consider two-dimensional nematic films with free interfaces subjected to competing boundary conditions. We focus on the coupling between the nematic director and the normal to a free interface, assuming the Rapini--Papoular form of the anchoring free energy~\cite{rapini:1969}. It is known~\cite{BB:1983} that the director orientation varies along the thickness $h$ of the film if $h>|K/W_1-K/W_2|=h_c$ where $K\simeq10^{-11}$~J/m is the Frank elastic modulus in the one-constant approximation~\cite{frank:1958} and $W_{1,2}$ are the anchoring strengths at two interfaces favouring orthogonal alignment of the director $\bn$. Here we show, that in the long-wavelength limit the very same thickness  $h_c$ corresponds to the onset of instability from a flat nematic film towards periodically modulated film independent of the surface tension $\gamma$. \rev{As we increase the thickness of the film, the difference in the anchoring angles at two interfaces favours the distortions of the director $\bn$, until it approaches the interfacial normal at $\pi/4$-angle. This  thickness corresponds to the upper instability threshold above which the film remains flat.} To the best of our knowledge these results, followed from the linear stability analysis, were not presented before in the literature. 

Although, thin nematic films were extensively studied experimentally and theoretically, see e.g.~\cite{LP:1995,cazabat:2011,benamar:2001,lin:2013,pla:2013} and references therein, the question about the interplay between a film profile and a possible director configuration was not fully addressed. Experimental observations~\cite{LP:1995,cazabat:2011} suggest that nematic LC form extended films or domains of various size and thickness, \rev{which} are {\it flat} on the length-scales larger than periodicity of elastic distortions (stripes) of the director. Hydrodynamic approach for spreading of nematic drops accounts for the time-evolution of the film profile, assuming a certain form of the director field~\cite{benamar:2001,lin:2013}. Therefore, it is of fundamental interest to study the thickness modulation caused by the director reorientation or {\it vice versa}  at the same length-scale.

The paper is organised as follows. Without referring to any particular experiment, first we consider a two-dimensional (2D) nematic film with competing boundary conditions and perform the linear stability analysis of a flat film. Next, we confine nematic LC to a radial geometry and explore the symmetry breaking of a ring induced by elastic distortions of the director.

\section*{2D thin nematic films}

\begin{figure}[tb]
\centering
\includegraphics[width=\linewidth]{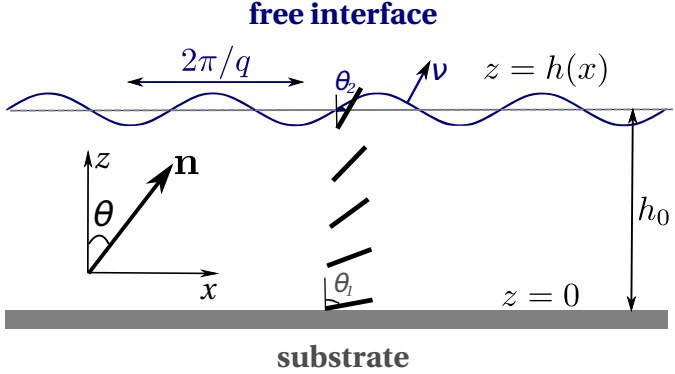}
\caption{Schematic representation of the 2D problem. Thin nematic film with director $\bn=\sin\theta\bi_x+\cos\theta\bi_z$, subjected to the competing boundary conditions. At the lower interface $z=0$ the anchoring is planar ($\bn\parallel \bi_x$ is equilibrium), at the free interface at $z=h(x)$ with the normal $\nu$  the  anchoring is homeotropic ($\bn\parallel \nu$ is equilibrium). Undulations of the free interface is chosen to be periodic with the wavelength $2\pi/q$.}
\label{fig:interf}
\end{figure}

We describe the  nematic LC by a vectorial order parameter $\bn$ ($|\bn|^2=1$), which can be decomposed into $\bn=\sin\theta \bi_x + \cos\theta\bi_z$ in the Cartesian $x$-$z$ coordinates. Here $\theta(x,z)$ is an angle between $\bn$ and the $z$-axis varying in space (see Fig.~\ref{fig:interf}). Another `slow' field of our model is the thickness of the nematic film $h(x)$. Then the normal to the free interface $\nu$ is given by $\nu=\nu_x\bi_x+\nu_z\bi_z=(-\p_x h \,\bi_x+\bi_z)/{\sqrt{1+(\p_xh)^2}}$.  Without restriction of generality we assume that: i) the nematic substrate interface $z=0$ is characterised by the planar anchoring, with equilibrium  configuration of $\bn$ along the $x$-axis, ii) at the free interface $z=h(x)$ we have homeotropic anchoring where the director $\bn$ tends to align along the normal $\nu$. Moreover, we assume that the anchoring contribution to the surface free energy per unit area can be written in a simple Rapini--Papoular form~\cite{rapini:1969} as
\begin{align}
\omega_{a_1}&=\bigg(\gamma_1+\frac{W_1}2 \cos^2\theta\bigg)\bigg|_{z=0}, \label{eq:oms1} \\
\omega_{a_2}&=\bigg(\gamma_2 + \frac{W_2}2 \big(1-(\bn\cdot\nu\big)^2)\bigg)\bigg|_{z=h(x)}=\notag\\&= \gamma_2 +\frac {W_2}2\frac{\sin^2\theta+\sin2\theta \p_x h+\cos^2\theta (\p_xh)^2 }{1+(\p_xh)^2}\bigg|_{z=h(x)},\label{eq:oms2}
\end{align}
where $\gamma_i$ is the isotropic surface tension and $W_i$ is the anchoring strength  at the nematic--substrate ($i=1$) and the free ($i=2$) interfaces. For a flat free interface ${\p_xh\equiv0}$  we recover $W_2 \sin^2\theta/2$, which is the usual Rapini--Papoular form  of the anchoring free energy. The competing boundary conditions force the director $\bn$ to vary along the thickness of the film, which cost additional bulk free energy per unit volume associated with elastic deformations
\be\label{eq:omb}
\omega_b = \frac K 2 |\nabla \bn|^2=\frac K2 |\nabla \theta|^2=\frac K2 \big((\p_x\theta)^2+(\p_z\theta)^2\big),
\ee
where $K$ is the Frank~\cite{frank:1958} elastic constant in the  one-constant approximation. For thick LC films $h\gg|K/W_2-K/W_1|\equiv h_c$  deformations in the bulk $\int dx\, dz\,\omega_b\sim K/h$ are not energetically expensive and the total surface energy $\int dx(\omega_{a_1}+\sqrt{1+(\p_xh)^2}\omega_{a_2})$ can be minimised without distortions of the free interface, thus the film remains flat, $\p_x h=0$. For thin nematic films $h\simeq h_c$, when the bulk contribution $\omega_b$ is of the same order as the surface free energy $\omega_{a_i}$, the situation is not clear. Spontaneous deformations of the film profile $\p_x h\neq0$ together with the in-plane distortions of $\bn$ may lower the total free energy. In the following we perform the linear stability analysis and derive the conditions for instability of a flat nematic film under competing anchoring terms.


The thickness $h$ of the domain is a free parameter adopted by our system~\cite{pla:2013}, playing the role of magnetic field responsible for instability in~\cite{degennes:1970}. We can formulate a variational problem for the director $\bn$. The vanishing of the first variation of the total free energy, given by the sum of \eqref{eq:oms1}--\eqref{eq:omb}, yields the Euler--Lagrange equation for $\theta$ and two natural boundary conditions $\big(\nu\cdot{\p \omega_b}/({\p\nabla\theta})\big)+{\p\omega_{a_i}}/{\p\theta}$ at $z=0$ and $z=h(x)$, such as
\begin{gather}
\p_{xx} \theta+\p_{zz} \theta=0,\label{eq:EL}\\[1ex]
K\p_z\theta|_{z=0}+W_1\sin\theta\cos\theta|_{z=0}=0,\label{eq:bc1}\\[1ex]
K\big(\nu_x\p_x\theta+\nu_z\p_z\theta\big)\big|_{z=h(x)} +W_2\big(\sin\theta\cos\theta\nu_z^2-\notag\\-\cos2\theta \nu_x\nu_z-\cos\theta\sin\theta \nu_x^2 \big)\big|_{z=h(x)}=0.\label{eq:bc2}
\end{gather}
Note that at the lower interface we have $\nu|_{z=0}=-\bi_z$.
We  are looking for solution of this system in the following form
\begin{subequations}
\label{eq:sol}
\begin{align}
\theta(x,z)&=\theta_0(z)+\ve\, \hat \theta(z){\cos (q x)}, \\
h(x)&=h_0\big(1+\ve \lambda\cos(q x)\big),
\end{align}
\end{subequations}
where $\ve\ll1$ is a small parameter and $q$ is the wavenumber of the periodic distortions along $x$-direction. This {\it ansatz} implies that the variation of the profile and the director happen at the same order and on the same length scale~$2\pi/q$. 
 
Substituting the form \eqref{eq:sol}  into \eqref{eq:EL} we get
\begin{alignat}{2}\label{eq:theta0}
O(1):& \qquad &\theta_0(z)&=\theta_1+(\theta_2-\theta_1) \frac z{h_0},\\
O(\ve):&\qquad &\hat\theta(z)&=A\sinh(q z)+B\cosh(qz).\label{eq:thetahat}
\end{alignat}
The equilibrium anchoring angles $\theta_{1,2}$ (see Fig.~\ref{fig:interf}) satisfy the boundary conditions \eqref{eq:bc1}, \eqref{eq:bc2} at $O(1)$ 
\begin{subequations}
\label{eq:bc0}
\begin{align}
\frac {L_1} {h_0}(\theta_2-\theta_1)+\sin\theta_1\cos\theta_1&=0,\qquad L_1=\frac K{W_1}\\
\frac {L_2} {h_0}(\theta_2-\theta_1)+\sin\theta_2\cos\theta_2&=0,\qquad L_2=\frac K{W_2}.
\end{align}
\end{subequations}
The next order contribution to the boundary conditions $O(\ve)$ establishes the connection between the amplitude $\lambda$ of the deformed film profile and the amplitude of the director modulation $\hat\theta(z)$~\eqref{eq:sol}, yielding
\begin{subequations}
\label{eq:AB}
\begin{align}
A&=\frac{ \lambda (\theta_2-\theta_1)\cos(2 \theta_1) \cos(2 \theta_2)}{ {\cal P}},\\[1ex]
B&=\frac{\lambda \xi_1 \chi (\theta_1 - \theta_2)  \cos(2 \theta_2)}{ {\cal P}}, 
\end{align}
\end{subequations}
where $\chi\equiv q h_0$ is the dimensionless wavenumber,  $\xi_1\equiv L_1/h_0$ and $\xi_2\equiv L_2/h_0$ are extrapolation anchoring lengths at two interfaces scaled by $h_0$, ${\cal P}\equiv\chi  \cosh\chi \big[ \xi_1 \break\cos(2 \theta_2)-\xi_2 \cos(2 \theta_1) \big]  + \sinh\chi \big[\xi_1 \xi_2 \chi^2 - \cos(2 \theta_1) \cos(2 \theta_2)\big]$. Immediately, from~\eqref{eq:AB} it follows that the periodic distortions of the director $\hat\theta(z)\neq0$ happen if and only if i) $\theta_1\neq \theta_2$, we have a hybrid aligned nematic (HAN) state, satisfying~\eqref{eq:bc0} and ii) $\lambda\neq 0$, a non-flat profile of the film is favoured. Note that in the long-wavelength limit $\chi\to0$ we get $B\to0$.

\begin{figure}[tb]
\centering
\includegraphics[width=\linewidth]{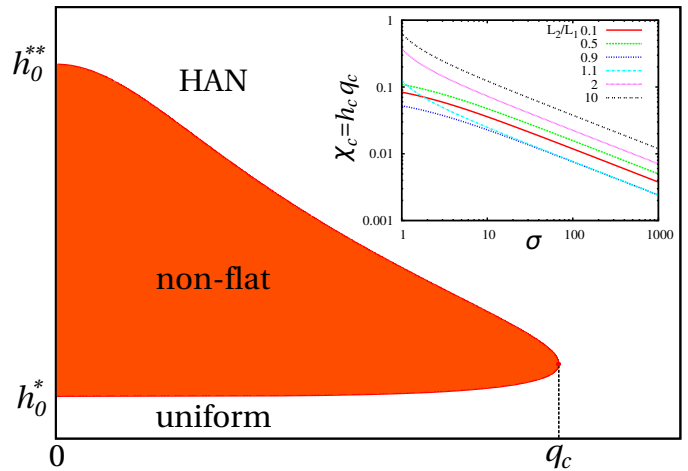}
\caption{The critical region in $h_0$--$q$ plane, when the film is non-flat $h_0^*<h_0<h_0^{**}$~\eqref{eq:hstar}. The film is flat with uniform nematic director for $h<h_0^{*}$ and hybrid aligned nematic (HAN) for $h>h_0^{**}$. Inset: the approximated critical wavenumber $\chi_c=q_ch_c$ ($h_0|_{q=q_c}\approx h_c=|L_2-L_1|$) as function of the surface tension  $\sigma=\gamma_2/W_2$ in logarithmic scale for different anchoring strengths $L_2/L_1\equiv W_1/W_2$.}
\label{fig:hstar}
\end{figure}

We are interested in a non-trivial solution to the problem, when the flat film becomes unstable towards periodic modulations with $\lambda\neq0$. To quantify this instability we use the equilibrium solutions~\eqref{eq:sol}--\eqref{eq:AB} and integrate directly the total free energy~(sum of \eqref{eq:oms1}--\eqref{eq:omb}) over the thickness $h(x)$ and the period $x\in[0,2\pi/q]$ with the help of {\it Mathematica} (see appendix~\eqref{eq:df2}). \rev{Below we focus on the critical threshold without solving the amplitude equations for $\lambda$, which would require the next higher order approximation of the total free energy.} \rev{Flat films are linearly stable if the free energy contribution quadratic in $\lambda$ is positive (${\cal F}^{(2)}>0$ \eqref{eq:df2} with $\theta_{1,2}$ satisfying~\eqref{eq:bc0}). Otherwise the film profile is unstable to periodic perturbations with non-zero amplitude $\lambda$ and wavenumber $q\neq0$. Thus there exists a non-trivial solution $\hat\theta(z)$~\eqref{eq:thetahat} extremising the free energy.} The result is summarised in Fig.~\ref{fig:hstar}, where we show the region of (in)stability of (non)flat films in $h_0$--$q$ plane. Independent of the surface tension $\gamma_2$, in the long-wavelength limit we find the elegant closed form, characterising the onset of this instability 
\begin{multline}\label{eq:hstar}
|L_1-L_2|\equiv h_0^*<h_0< h_0^{**}\equiv\\[1ex]\equiv \left\{\begin{aligned}&L_2\bigg(\frac \pi2-\sin^{-1}\frac{L_1}{L_2}),\quad \mbox{for }L_1<L_2,\\ &L_1\bigg(\frac \pi2-\sin^{-1}\frac{L_2}{L_1}),\quad \mbox{for }L_1>L_2.\end{aligned}\right.
\end{multline}
The lower threshold $h_0^*=|L_1-L_2|=h_c$ coincides with the Barbero--Barberi critical thickness~\cite{BB:1983}, characterising the transition between the uniform nematic state (with $\theta_1=\theta_2$) towards the HAN state ($\theta_1\neq\theta_2$). The upper instability threshold~$h_0^{**}$ corresponds to the thickness when $\theta_2=\pi/4$ ($L_1<L_2$) or  $\theta_1=\pi/4$ ($L_1>L_2$), which follows from the boundary conditions for the equilibrium angles $\theta_{1,2}$~\eqref{eq:bc0}. \rev{Note that the critical anchoring angle of $\pi/4$ is not universal, rather it is a consequence of the analysis with the assumed Rapini--Papoular form for the surface energy~\cite{rapini:1969} and the one elastic constant approximation.}

In the inset of Fig.~\ref{fig:hstar} we plot the estimated critical wavenumber $q_c$ as function of $\sigma=\gamma_2/W_2$, characterising the relative contribution of the capillary versus anchoring forces at the free interface. For different ratios of the anchoring strength $L_2/L_1\equiv W_1/W_2$, the modulation of the film profile and the nematic director $q_c$~\eqref{eq:sol} strongly depends on the surface tension $\sigma$.  Since the critical thickness $h_0|_{q=q_c}\simeq h_c$, the analysis is significantly simplified and 
the critical wavelength  is roughly proportional to the square root of the surface tension, $\lambda_c\simeq h_c/\chi_c\simeq f({L_2}/{L_1}) h_c\sqrt{\sigma}$ (see appendix). Then for $h_c$ of the order of $1~\mu$m and $\sigma=1000$ we get $\lambda_c$ of the order of hundreds of micrometers. We find a good agreement by comparing approximated $\chi_c$ in the inset of Fig.~\ref{fig:hstar} with the exact curves of the free energy at $O(\ve^2)$ written explicitly in the appendix.

\section*{2D nematic confined in a ring}
\begin{figure}[tb]
\centering
\includegraphics[width=\linewidth]{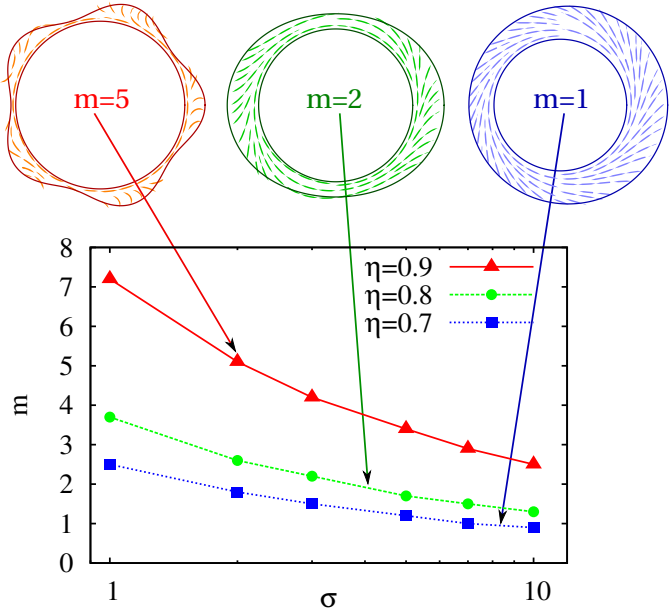}
\caption{The critical `wavenumber' $m_c$ as function of $\sigma=\gamma_2/W_2$ for different ratios of the inner and outer radii, $\eta=R_1/R_0$. We assume an infinitely strong anchoring $L_1\to 0$ at the inner radius $r=R_1$. Top: the equilibrium solutions at $m_c=5,2,1$~\eqref{eq:solp}. To sketch the director configurations we choose the amplitude $\ve=0.2$ and $\lambda=0.4$, which cannot be found from the linear stability analysis.}
\label{fig:m2s}
\end{figure}

In this section we confine nematic liquid crystal to an annular geometry with $r\in[R_1,R_2]$, $\vp\in[0,2\pi]$ and we aim to study the instabilities of the director configuration under competing boundary conditions. The inner boundary $r=R_1$ is fixed and favours the tangential alignment of the director  $\bn=\bi_\vp$. The outer boundary $r=R_2(\vp)$ is free and the corresponding anchoring energy is minimised when $\bn$ coincides with the normal $\nu=\nu_r\bi_r+\nu_\vp\bi_\vp=(R_2\bi_r+\p_\vp R_2\bi_\vp)/\sqrt{R_2^2+(\p_\vp R_2)^2}$. We are interested in the equilibrium solutions of the director $\bn=\cos\alpha\,\bi_r+\sin\alpha\,\bi_\vp$ and plausible deformations of the free interface, within the {\it ansatz} similar to~\eqref{eq:sol}
\begin{subequations}
\label{eq:solp}
\begin{align}
R_2(\vp)&=R_0\big(1+\ve \lambda\cos(m\vp)\big),\\
\alpha(r,\vp)&=\alpha_0(r)+\ve\, \hat \alpha(r){\cos (m \vp)}. 
\end{align}
\end{subequations}
Here  $m$ is an integer due to the closure conditions $\alpha(\vp+2\pi)=\alpha(\vp)$ and $R_2(\vp+2\pi)=R_2(\vp)$. The equilibrium solutions $\alpha_0(r)$ and $\hat\alpha(r)$  satisfy the Euler--Lagrange equation
\be
\p_r(r\p_r \alpha)+\frac 1 r \p_{\vp\vp} \alpha=0,
\ee
and read as $\alpha_0(r)=\alpha_2 + (\alpha_1 - \alpha_2) \log(r/R_0)/ \log(\eta)$, where $\eta=R_1/R_0<1$ and $\hat\alpha(r)=A \sinh(m \log(r/R_0))+B\cosh(m\log(r/R_0))$. From the boundary conditions 
\begin{subequations}
\label{eq:bc0p}
\begin{align}
\frac {2(\alpha_1-\alpha_2)} {\eta\log\eta}+\frac{R_0}{L_1}\sin2\alpha_1&=0,\\
\frac {2(\alpha_1-\alpha_2)}{\log\eta}+\frac{R_0}{L_2}\sin2\alpha_2&=0,
\end{align}
\end{subequations}
we find the equilibrium anchoring angles $\alpha_{1,2}$, similar to $\theta_{1,2}$ in the previous section. 
Also we deduce that the director configuration with $\alpha_1=\alpha_2$ is stable when $R_0<|(L_1-\eta L_2)/(\eta\log\eta)|$. \rev{This relationship defines implicitly (but uniquely) the critical outer radius $R_c$, analogous to the Barbero--Barberi critical thickness~\cite{BB:1983}, given the anchoring strengths $L_{1,2}$ and the size of the inner radius~$R_1$.} As discussed in the previous section, for $R_0>R_c$ one expects an instability towards deformed  profile~\eqref{eq:solp} with $m\neq0$, $\lambda\neq0$ and distorted nematic director, according to~\eqref{eq:solp}. \rev{Assuming  for simplicity
a strong planar anchoring with $L_1\to 0$ at $r=R_1$,  we find the following expression for the critical radius $R_c=L_2/{\cal W}(L_2/R_1)$ with the Lambert ${\cal W}$ special function.} In Fig.~\ref{fig:m2s} for $R_0>R_c$ we show the critical number of periods $m_c$ as function of $\sigma=\gamma_2/W_2$~\eqref{eq:oms2}. The symmetry breaking of an outer circle (with $m=0$) towards structures with $m$-fold rotational symmetry happens for $\sigma\simeq 1$, when  the anchoring strength  $W_2$ is of the same order of magnitude as the surface tension~$\gamma_2$. The number of folds $m_c$ increases with $\eta$ and the instability happens at bigger radii. \rev{A similar increase of the wavenumber, when thickness of the nematic confined to the annular geometry decreases, was found in~\cite{gaetano:2010}. In that case, however, the driving force for the periodic distortions of ${\bf n}$ is  the anisotropy of elastic constants and presence of an electric field.} 


\section*{Concluding remarks}

The presented phenomenological model explores a possible interplay between the nematic director configuration  and the shape of free interface in planar and radial geometries. The main prediction of the model is the existence of parameter range where the interesting phenomena related to interface instability may occur. In particular, we calculated the critical wavelength of interface undulations, which can be potentially tested in experiments. To achieve a better quantitative comparison, further improvements and extensions of this simplified model are required, for example,  i)~accounting for the anisotropy of splay $K_1$ and bend $K_3$ Frank elastic moduli in 2D; ii)~performing a (weakly) non-linear analysis to find an amplitude $\lambda$ of interface distortions. Solving a fully non-linear problem is truly challenging. We believe that the presented findings would spark experimental interest, which in turn will guide further developments of the theory, where the shape and structure of liquid crystalline materials are intertwined. The authors of~\cite{lavrent:2013} have recently studied the formation of cusps in nematic tactoids and demonstrated the role of elastic anisotropy on the resulting non-circular shape. Considering radial geometries could be relevant to characterise non-trivial morphologies of biological systems (e.g.~\cite{benamar:2011}) where the effects of surface tension and anisotropic elastic forces acting on different length-scales are not well understood.

\acknowledgement{I am indebted to T.-S. Lin and U. Thiele for collaboration and helpful discussions. I am grateful to O. D. Lavrentovich for critical reading of the manuscript. Part of this work was done at the Isaac Newton Institute for Mathematical Sciences in Cambridge, which I thank for its hospitality and acknowledge financial support.}
\appendix
 
\section{Taylor expansion of the free energy}
The next order contribution to the free energy is

\begin{widetext}%
\begin{multline}\label{eq:df2}
{\cal F}^{(2)}\propto \frac{K \lambda^2\chi}{32 {\cal P} h_0 \xi_2 } 
\Big\{ \cosh\chi \big[8\xi_2(\theta_1-\theta_2)^2\cos(2\theta_1)\cos(2\theta_2) +2 \chi^2 \big(1 + 4 \sigma + 3 \cos(2 \theta_2)\big)\big(\xi_1\cos(2 \theta_2)-\xi_2\cos(2\theta_1)\big)+\\+
4 \cos(2\theta_1)(\theta_2 - \theta_1) \sin(4 \theta_2)\big]+  2\chi  \sinh\chi\big[
\big(1 + 4 \sigma + 3 \cos(2 \theta_2)\big)\big(\xi_1 \xi_2 \chi^2 -\cos(2 \theta_1)\cos(2 \theta_2)\big)+\\
 +  4(\theta_1 - \theta_2) \xi_1 \cos(2 \theta_2)\big(\sin(2\theta_2)+  \xi_2 (\theta_2 -\theta_1)\big) \big] \Big\}.
\end{multline}
\end{widetext}
\rev{ According to the Landau theory of the second-order phase transitions we may treat $\lambda$ as an order parameter. Then a flat film with $\lambda=0$ is the equilibrium solution if ${\cal F}^{(2)}>0$, otherwise an instability towards a non-flat film occurs (see Fig.~\ref{fig:hstar}).} Although, we can explore the parameter space directly, using~\eqref{eq:df2}, we are interested in the approximate behaviour of ${\cal F}^{(2)}$ in several limiting cases: 
\begin{itemize}
\item[i)] the long-wavelength limit ($\chi\to 0$) 
$$
{\cal F}^{(2)}\propto\frac{K\lambda^2\Delta\theta \cos(2 \theta_1) \cos(2 \theta_2) \big[h_0 \sin(2 \theta_2) -L_2 \Delta\theta\big]}{4 h_0^2 L_2 \big[\cos(2 \theta_1) (L_2+h_0 \cos(2 \theta_2))-L_1 \cos(2 \theta_2)\big]},
$$
where $\Delta\theta\equiv \theta_1-\theta_2$, for a given thickness $h_0$ the angles $\theta_{1,2}$ can be computed from the boundary conditions~\eqref{eq:bc0}.
\item[ii)] In the vicinity of the lower threshold $h_0^*=h_c=|L_1-L_2|$, we assume  $h_0=h_c(1+\delta h)$ and $\theta_{1,2}=\pi/2-\delta\theta_{1,2}$ (or $\theta_{1,2}=\delta\theta_{1,2}$ for $L_1>L_2$), where $\delta h, \delta\theta_{1,2}\ll 1$ are small perturbations, related through~\eqref{eq:bc0}, so that 
$$
\delta\theta_1= \sqrt{\frac {3L_1^2\, \delta h}{2(L_1^2+L_1 L_2+L_2^2)}}, \quad \delta\theta_2=\frac{L_2}{L_1}\delta\theta_1.
$$
Replacing the above relations into~\eqref{eq:df2} we get ${\cal F}^{(2)}\propto a_0+a_1\delta h+a_2(\delta h)^2+O(\delta h^3)$ and thus the critical point is determined by solving the system of equations $4 a_0a_2-a_1^2=0$ and $\delta h=-a_1/(2a_2)$. The critical wavenumber $q_c$ is shown in the inset of Fig.~\ref{fig:hstar}. The correction to the critical thickness is negligible: $\max(\delta h)\simeq0.05$ at $\sigma=1$, and $\max(\delta h)\simeq0.001$ at $\sigma=100$. Therefore the assumption of $h_0|_{q=q_c}\simeq h_c$, made to plot $\chi_c$, is well-justified.  
\end{itemize}


\end{document}